\documentclass[a4paper,10pt]{article}

\usepackage[utf8]{inputenc}
\usepackage{xspace}
\usepackage{tikz}
\usepackage[labelformat=simple]{subcaption}

\usepackage{amsthm}

\newtheorem{theorem}{Theorem}
\newtheorem{definition}[theorem]{Definition}
\newenvironment{proofof}[1]{\bigskip \noindent {\bf Proof of #1:}\ }{\qed\par\vskip 4mm\par}

\usetikzlibrary{calc}

\newcommand{\nodesize}{6pt}

\tikzstyle{every picture}+=[
node/.style={circle, minimum size=\nodesize, draw=black, inner sep=0,fill=white},       
blacknode/.style={circle, minimum size=\nodesize, draw=black, inner sep=0,fill=black},
empty/.style={draw=none,fill=none},
curvedLine/.style={decorate,decoration={snake,amplitude=.5mm}}
]

\tikzstyle{every node}=[font=\footnotesize]

\newcommand{\nodewithlabelleft}[3]{
	\node[node](#1) at (#2) {};		
	\node[left] at (#1.west) {#3};
}
\newcommand{\nodewithlabelright}[3]{
	\node[node](#1) at (#2) {};		
	\node[right] at (#1.east) {#3};
}
\newcommand{\nodewithlabelabove}[3]{
	\node[node](#1) at (#2) {};		
	\node[above] at (#1.north) {#3};
}
\newcommand{\nodewithlabelbelow}[3]{
	\node[node](#1) at (#2) {};		
	\node[below] at (#1.south) {#3};
}
\newcommand{\curvedconnection}[3]{
	\draw (#1) .. controls ($ (#1)!.1!(#2)+(0,#3) $) and ($ (#1)!.9!(#2)+(0,#3) $) .. (#2);
}

\newcommand{\minLAP}{\textsc{minLA}\xspace}
\newcommand{\pminLAP}{\textsc{planar minLA}\xspace}
\newcommand{\ARR}{\pi}
\newcommand{\ARRP}{\ARR_p\xspace}
\newcommand{\minpa}{minimum planar arrangement\xspace}
\newcommand{\minpas}{minimum planar arrangements\xspace}

\title{The \pminLAP is different from the \minLAP}
\author{Alexander Setzer}

\begin{document}

\maketitle

\begin{abstract}
In various research papers, such as \cite{bib:petit2011}, one will find the claim that the \minLAP is optimally solvable on outerplanar graphs, with a reference to \cite{bib:frederickson1988}.
However, the problem solved in that publication, which we refer to as the \pminLAP, is different from the \minLAP, as we show in this article.
\end{abstract}

In constrast to the minimum linear arrangement problem (\minLAP), the planar minimum linear arrangement problem (\pminLAP) poses an additional restriction on the arrangements:
It must be possible to draw all edges of the input graph $G$ such that they ``run above'' the nodes and do not intersect.
More formally:
\begin{definition}[Crossing edges]
 Let $G=(V,E)$ be a graph and let $\ARR$ be a linear arrangement of $G$.
	Two distinct edges $\{u,v\},\{x,y\}\in E$ \emph{cross} if: $\ARR(u) < \ARR(x) < \ARR(v) < \ARR(y)$.
\end{definition}

\begin{definition}[Minimum planar arrangement]
 A \emph{\minpa} of an input graph $G=(V,E)$ is a mapping $\ARR: V \rightarrow \{1, \dots, n\}$ such that no two edges of $G$ cross in $\ARR$.
\end{definition}

We prove that optimal solutions of the \pminLAP are different from optimal solutions of the \minLAP by presenting a counterexample.
That is, we give an example graph whose corresponding minimum linear arrangement yields a smaller cost than all possible \minpas.

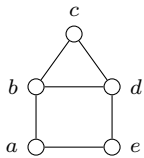
\begin{figure}[htb]\centering
	\begin{tikzpicture}
	\nodewithlabelleft{a}{0,0}{$a$}
	\nodewithlabelleft{b}{0,0.8}{$b$}
	\nodewithlabelabove{c}{0.5,1.5}{$c$}
	\nodewithlabelright{d}{1,0.8}{$d$}
	\nodewithlabelright{e}{1,0}{$e$}
		\draw (a) -- (b) -- (c) -- (d) -- (e) -- (a);
		\draw (b) -- (d);

	\end{tikzpicture}
	\caption{Input graph used to prove the counterexample.}\label{appendix:fig:input}
\end{figure}

The input graph $G=(V,E)$ we use is given by Figure~\ref{appendix:fig:input}.

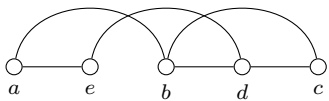
\begin{figure}[htb]\centering
	\begin{tikzpicture}
	\nodewithlabelbelow{a}{0,0}{$a$}
	\nodewithlabelbelow{e}{1,0}{$e$}
	\nodewithlabelbelow{b}{2,0}{$b$}
	\nodewithlabelbelow{d}{3,0}{$d$}
	\nodewithlabelbelow{c}{4,0}{$c$}

 		\curvedconnection{a}{b}{1.0}			
		\curvedconnection{b}{c}{1.0}	
		\draw (c) -- (d);		 		
		\curvedconnection{d}{e}{1.0}
		\draw (e) -- (a);			
		\draw (b) -- (d);		
	\end{tikzpicture}
	\caption{Arrangement $\ARR_1$ of $G$. The cost of $\ARR_1$ is $9$.}\label{appendix:fig:arrangement}
\end{figure}

We claim that the arrangement $\ARR_1$ given by Figure~\ref{appendix:fig:arrangement} yields a lower cost than any \minpa.

\begin{theorem}\label{appendix:thm:minpa}
 Any \minpa of $G$ has a cost stricly larger than $9$.
\end{theorem}

In order to prove this, we determine all \minpas of $G$ (which are exactly five, plus their symmetric counterparts, as we will see).
For this, we need the following terminology, which is taken from \cite{bib:frederickson1988}:
\begin{definition}[Dominating edge]
 Let $G=(V,E)$ be a graph and let $\ARRP$ be a \minpa of $G$.
	An edge $\{x,y\}$ \emph{dominates} an edge $\{u,v\}$, if $\{x,y\} \neq \{u,v\}$ and $\ARRP(u) \leq \ARRP(x) < \ARRP(y) \leq \ARRP(v)$.
\end{definition}

Provided with this definition, we are ready to prove Theorem \ref{appendix:thm:minpa}:

\begin{proofof}{Theorem \ref{appendix:thm:minpa}}
	The idea of this proof is to identify all possible \minpas and to show that the cost of each such arrangement is greater than $9$.
	Instead of testing all permutations of the nodes in $G$, we find conditions of \minpas for $G$ and then only consider graphs that fulfill these conditions.

	Let $\ARR_p$ be a \minpa of $G$.
	Moreover, let $E_C$ be the set of edges on the cycle $(a,b,c,d,e,a)$ in $G$.
	We first show:
	\begin{enumerate}
		\item For each edge $\{u,v\} \in E_C$, one of the following is true:
			\begin{enumerate}
			\item $\{u,v\}$ dominates all other edges in $\ARR_p$
			\item $u$ and $v$ are neighbors in $\ARR_p$
			\end{enumerate}
		\item There is exactly one edge $\{u,v\} \in E_C$ that dominates all other edges.
	\end{enumerate}

	For the first claim, assume that for an arbitrary edge $\{u,v\} \in E_C$, none of the two cases is true, i.e., neither does $\{u,v\}$ dominate all other edges, nor are $u$ and $v$ neighbors in $\ARR_p$.
	This implies there are exactly one or two nodes between $u$ and $v$.
	However, observe that for each of the edges $\{u,v\} \in E_C$, the subgraph induced by $V \setminus \{u,v\}$ contains a path of length three.
	Independent of which single or two nodes we place between $u$ and $v$, this path crosses $\{u,v\}$.
	Therefore, this is not possible without violating the constraints of a \minpa and the first claim is proven.

	For the second claim, assume for contradiction that the claim is not true, i.e. there is no edge  $\{u,v\} \in E_C$ that dominates all other edges.
	This implies, by the first claim, for each edge $\{x,y\} \in E_C$ that $x$ and $y$ are neighbors in $\ARR_p$.
	However, since the edges in $E_C$ form a cycle, this is not possible for all edges.
	This completes the proof of the second claim.

	Now, let $\{u,v\} \in E_C$ be the edge that dominates all other edges in $\ARR_p$.
	The first claim implies that for the edges $\{x,y\} \in E_C \setminus \{u,v\}$, $x$ and $y$ must be neighbors in $\ARR_p$.
	Since the edges in $E_C\setminus \{u,v\}$ form a path from $u$ to $v$, this uniquely defines the order of the other nodes in $\ARR_p$.

	Provided with this, we can derive all possible \minpas by selecting an edge $\{u,v\} \in E_C$, putting $u$ and $v$ at the positions $1$ and $5$, and placing the other nodes such that for $\{x,y\} \in E_C \setminus \{u,v\}$, $x$ and $y$ are neighbors in the arrangement.
	This yields five possible \minpas (except for symmetry):
\begin{figure}[Hhtb]\centering
	\begin{subfigure}{125pt}\centering
		\begin{tikzpicture}
		\nodewithlabelbelow{a}{0,0}{$a$}
		\nodewithlabelbelow{e}{1,0}{$e$}
		\nodewithlabelbelow{d}{2,0}{$d$}
		\nodewithlabelbelow{c}{3,0}{$c$}
		\nodewithlabelbelow{b}{4,0}{$b$}

			\curvedconnection{a}{b}{1.5}			
			\draw (b) -- (c);		 
			\draw (c) -- (d);		 		
			\draw (d) -- (e);		 
			\draw (e) -- (a);			
			\curvedconnection{b}{d}{0.9}	
		\end{tikzpicture}
		\caption{Arrangement with $a$ and $b$ at the outmost positions. Its cost is $10$.}
	\end{subfigure}
	\hfill
	\begin{subfigure}{125pt}\centering
		\begin{tikzpicture}
		\nodewithlabelbelow{b}{0,0}{$b$}
		\nodewithlabelbelow{a}{1,0}{$a$}
		\nodewithlabelbelow{e}{2,0}{$e$}
		\nodewithlabelbelow{d}{3,0}{$d$}
		\nodewithlabelbelow{c}{4,0}{$c$}

			\curvedconnection{b}{c}{1.5}			
			\draw (a) -- (b);		 
			\draw (c) -- (d);		 		
			\draw (d) -- (e);		 
			\draw (e) -- (a);			
			\curvedconnection{b}{d}{0.9}	
		\end{tikzpicture}
		\caption{Arrangement with $b$ and $c$ at the outmost positions. Its cost is $11$.}
	\end{subfigure}
	\hfill
	\begin{subfigure}{125pt}\centering
		\begin{tikzpicture}
		\nodewithlabelbelow{c}{0,0}{$c$}
		\nodewithlabelbelow{b}{1,0}{$b$}
		\nodewithlabelbelow{a}{2,0}{$a$}
		\nodewithlabelbelow{e}{3,0}{$e$}
		\nodewithlabelbelow{d}{4,0}{$d$}

			\curvedconnection{c}{d}{1.5}			
			\draw (a) -- (b);		 
			\draw (b) -- (c);		 		
			\draw (d) -- (e);		 
			\draw (e) -- (a);			
			\curvedconnection{b}{d}{0.9}	
		\end{tikzpicture}
		\caption{Arrangement with $c$ and $d$ at the outmost positions. Its cost is $11$.}
	\end{subfigure}
	\hfill
	\begin{subfigure}{125pt}\centering
		\begin{tikzpicture}
		\nodewithlabelbelow{d}{0,0}{$d$}
		\nodewithlabelbelow{c}{1,0}{$c$}
		\nodewithlabelbelow{b}{2,0}{$b$}
		\nodewithlabelbelow{a}{3,0}{$a$}
		\nodewithlabelbelow{e}{4,0}{$e$}

			\curvedconnection{d}{e}{1.5}			
			\draw (a) -- (b);		 
			\draw (b) -- (c);		 		
			\draw (c) -- (d);		 
			\draw (e) -- (a);			
			\curvedconnection{b}{d}{0.9}	
		\end{tikzpicture}
		\caption{Arrangement with $d$ and $e$ at the outmost positions. Its cost is $10$.}
	\end{subfigure}
	\hfill
	\begin{subfigure}{125pt}\centering
		\begin{tikzpicture}
		\nodewithlabelbelow{a}{0,0}{$a$}
		\nodewithlabelbelow{b}{1,0}{$b$}
		\nodewithlabelbelow{c}{2,0}{$c$}
		\nodewithlabelbelow{d}{3,0}{$d$}
		\nodewithlabelbelow{e}{4,0}{$e$}

			\curvedconnection{a}{e}{1.5}			
			\draw (a) -- (b);		 
			\draw (b) -- (c);		 		
			\draw (c) -- (d);		 
			\draw (e) -- (d);			
			\curvedconnection{b}{d}{0.9}	
		\end{tikzpicture}
		\caption{Arrangement with $a$ and $e$ at the outmost positions. Its cost is $10$.}
	\end{subfigure}
	\hfill

	\caption{The possible \minpas of $G$.}
\end{figure}
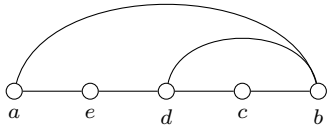
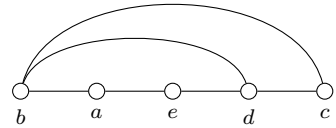
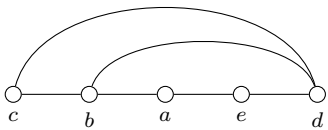
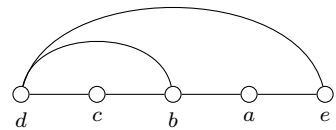
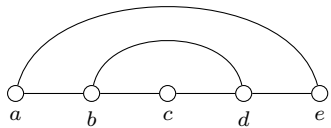

	All these arrangements have a cost of more than $9$.
\end{proofof}

    \bibliographystyle{plain}
    \bibliography{literature}

\end{document}